\documentclass{ws-procs9x6}
\usepackage{epsfig}
\begin{document}
\title{Spin Physics at RHIC - a Theoretical Overview}
\author{Marco Stratmann}
\address{Inst.~for Theor.~Physics,
Univ.~of Regensburg, D-93040 Regensburg, Germany \\
E-mail: marco.stratmann@physik.uni-regensburg.de}
\maketitle
%
\abstracts{
%
We review how RHIC is expected to deepen our understanding of the 
spin structure of longitudinally and transversely polarized nucleons.
After briefly outlining the current status of spin-dependent
parton densities and pointing out open questions, we focus on theoretical calculations 
and predictions relevant for the RHIC spin program.
Estimates of the expected statistical accuracy for such measurements are presented,
taking into account the acceptance of the RHIC detectors.}
%
\section{Lessons from (Un)polarized DIS}
\begin{figure}[th]
\vspace*{-0.5cm}
\centerline{\epsfxsize=0.63\textwidth\epsfbox{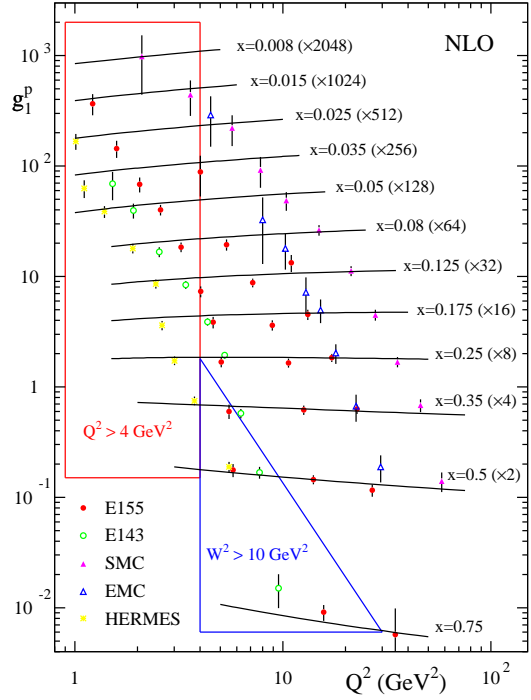}}   
\caption{Available information on $g_1(x,Q^2)$ as collected by 
fixed-target experiments$^1$ compared to results of a NLO QCD fit (solid lines).
The indicated rectangular and triangular regions contain data
which would not pass kinematical cuts of $Q^2>4\,\mathrm{GeV}^2$ 
and $W^2>10\,\mathrm{GeV}^2$, respectively, typically imposed in all fits
to unpolarized DIS data. \label{fig:fig1}}
\end{figure}

%
Before reviewing the prospects for spin physics at the BNL-RHIC we briefly
turn to longitudinally polarized deep-inelastic scattering (DIS)
and what we have learned from twenty years of beautiful data\cite{ref:expdata}.
Figure~\ref{fig:fig1} compares the available information on the DIS structure function
$g_1(x,Q^2)$ to results of a typical next-to-leading order (NLO) QCD fit.
From such types of analyses a pretty good knowledge of certain 
combinations of different quark flavors has emerged, and it became clear 
that quarks contribute only a small fraction to the proton's spin.
However, there is still considerable lack of knowledge regarding the
polarized gluon density $\Delta g$, which is basically unconstrained
by present data, the separation of quark and antiquark densities and of different
flavors, and the orbital angular momentum of quarks and gluons inside a nucleon.
In addition, spin effects with transverse polarization at the leading-twist level,
the so-called `transversity' densities, have not been measured at all. 
With the exception of orbital angular momentum RHIC can address all of these questions as will be
demonstrated in the following\cite{ref:rhicreport}.

There is also an important difficulty when analyzing polarized DIS data in terms
of spin-dependent parton densities: compared to the unpolarized case
the presently available kinematical coverage in $x$ and $Q^2$ and the statistical
precision of polarized DIS data are much more limited\cite{ref:expdata}. 
As a consequence, one is forced to include data into the fits from $(x,Q^2)$-regions where
fits of unpolarized leading-twist parton densities start to break down, see Fig.~\ref{fig:fig1}.
Data from RHIC, taken at `resolution' scales $Q^2$ where perturbative QCD and 
the leading-twist approximation are supposed to work, can shed light on the 
possible size of unwanted higher-twist contributions in presently available sets of 
polarized parton distributions. 

\section{Spin Physics at RHIC with Longitudinal Polarization}
%
\subsection{Prerequisites}
\begin{figure}[ht]
\vspace*{-0.2cm}
\begin{center}
\epsfig{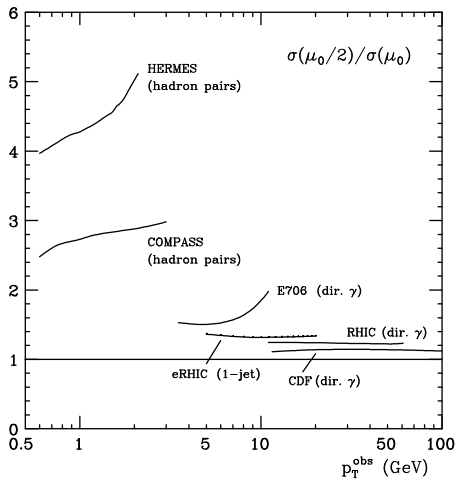}
\end{center}
\vspace*{-0.2cm}
\caption{Typical factorization scale dependence for various processes 
and experiments as a function of $p_T$.
Shown is the cross section ratio for two choices of scale,
$p_T$ and $p_T/2$. \label{fig:fig2}}
\vspace*{-0.4cm}
\end{figure}

%
The QCD-improved parton model has been successfully applied to many high energy 
scattering processes. The predictive power of perturbative QCD follows from the 
universality of the parton distributions and fragmentation functions. Once extracted
from data they can be used to make definite predictions for other processes.
This property is based on the factorization theorem where a cross section is written 
as a convolution of perturbatively calculable partonic hard scattering coefficients 
$d\hat{\sigma}_{ab}^c$ and appropriate parton densities $f_{a,b}$ and/or fragmentation
functions $D_c^H$. To be specific, let us consider the inclusive production of a hadron $H$,
e.g., a pion, in unpolarized proton-proton collisions:
\begin{equation}
\label{eq:eq1}
\frac{d\sigma^H}{d\Gamma} = \sum_{abc} \int dx_a\, dx_b\,dz\, f_a(x_a,\mu)\,
f_b(x_b,\mu)\,\frac{d\hat{\sigma}_{ab}^c}{d\Gamma}(x_a,x_b,z,{\Gamma},\mu)\,
D_c^H(z,\mu)\;.
\end{equation}
Here, $\Gamma$ stands for any appropriate set of kinematical variables like
the transverse momentum $p_T$ and/or rapidity $y$ of the observed hadron. 
The functions $f_{a,b}$ and $D_c^H$ embody non-perturbative physics. 
However, once they are known at some initial scale $\mu_0$, their scale $\mu$-dependence 
is calculable perturbatively via a set of evolution equations.
The factorization scale $\mu$, introduced on the r.h.s.\ of (\ref{eq:eq1}), separates 
long- and short-distance phenomena. $\mu$ is completely arbitrary but usually chosen to be 
of the order of the scale characterizing the hard interaction, for instance $p_T$. 
Since the l.h.s.\ of (\ref{eq:eq1}) has to be independent of $\mu$
(and other theoretical conventions), any residual
dependence of the r.h.s.\ on the actual choice of $\mu$ gives an indication of how well
the theoretical calculation is under control and can be trusted.
In particular, leading order (LO) estimates suffer from a strong, uncontrollable 
scale dependence and hence are not sufficient for comparing theory with data.
Figure~\ref{fig:fig2} shows a typical factorization scale dependence for various processes 
and experiments as a function of $p_T$. Clearly, the situation
is only acceptable at collider experiments where one can access
$p_T\gtrsim 5\,\mathrm{GeV}\ll \sqrt{S}/2=p_T^{\max}$ 
($\sqrt{S}$ is the available c.m.s.\ energy). 
$p_T$ values of about 1-2~GeV, accessible at fixed-target experiments, 
are {\em not} sufficient to provide a large enough hard scale to perform 
perturbative calculations reliably. For simplicity we have not distinguished between 
renormalization and initial/final-state factorization scales 
in (\ref{eq:eq1}) which can be chosen differently.

So far we have neglected the spin information contained in parton densities (and
fragmentation functions). Eq.~(\ref{eq:eq1}) can be easily extended to incorporate 
polarization by replacing all unpolarized quantities by their spin-dependent
counterparts, like, for instance, $f_{a,b} \to \Delta f_{a,b}$ and
$d\hat{\sigma}_{ab}^c/ d\Gamma \to d\Delta\hat{\sigma}_{ab}^c/d\Gamma$.
If a hard-scattering process with incoming protons having definite spin orientations
is studied, as at RHIC, one gains access to the spin distributions of quarks and
gluons in a longitudinally (or transversely) polarized proton.
In practice, spin experiments measure not the polarized cross section,
$d\Delta \sigma/d\Gamma$, itself, but the
spin asymmetry, which is given by the ratio of the polarized and unpolarized cross
sections, e.g., for our example, Eq.~(\ref{eq:eq1}), it reads
\begin{equation}
\label{eq:eq2}
A_{\mathrm{LL}}^H\equiv \frac{d\Delta \sigma^H/d\Gamma}{d\sigma^H/d\Gamma}\;\;\;.
\end{equation}
To denote the type of polarization of the colliding hadrons in (\ref{eq:eq2}) we use 
the subscripts `$L$=longitudinal' and `$T$=transverse'. At RHIC one can also study 
doubly transverse spin asymmetries, $A_{\mathrm{TT}}$, and single spin
asymmetries $A_{\mathrm{L}}$, $A_{\mathrm{T}}$ (the latter is often 
called $A_{\mathrm{N}}$) where only one of the protons is polarized.

\subsection{Accessing $\Delta g$}
The main thrust of the RHIC spin program\cite{ref:rhicreport} is to pin down the so far elusive
gluon helicity distributions $\Delta g(x,\mu)$. The strength of RHIC is the possibility
to probe $\Delta g(x,\mu)$ in a variety of hard processes\cite{ref:rhicreport},
in each case at sufficiently large $p_T$ where perturbative QCD is expected to work. 
This not only allows to determine the $x$-shape of $\Delta g(x,\mu)$ for 
$x\gtrsim 0.01$ but also verifies the universality property of polarized parton 
densities for the first time. In the following we review the status of theoretical
calculations for processes sensitive to $\Delta g$, experimental aspects can be 
found, e.g., in\cite{ref:matthias}.

The `classical' tool for determining the gluon density is high-$p_T$ prompt photon
production due to the dominance of the LO Compton process, $qg \to \gamma q$.
Exploiting this feature, both RHIC experiments, PHENIX and STAR, intend to use this process
for a measurement of $\Delta g$. 
Apart from `direct' mechanisms like $qg \to \gamma q$, the photon can also be
produced by a parton, scattered or created in a hard QCD reaction, which {\em fragments}
into the photon. Such a contribution naturally arises in a QCD calculation from the
necessity of factorizing final-state collinear singularities into a photon
fragmentation function $D_c^\gamma$. However, since photons produced through fragmentation
are always accompanied by hadronic debris, an `isolation cut' imposed on the photon signal
in experiment, e.g., a `cone', strongly reduces such contributions to the cross section.

\begin{figure}[th]
\vspace*{-0.3cm}
\centerline{\epsfxsize=3.1in\epsfbox{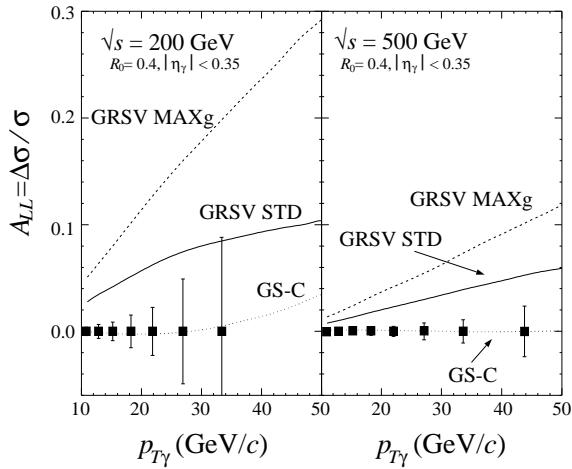}}   
\caption{$A_{\mathrm{LL}}$ for prompt photon production in NLO QCD as a
function of $p_T$ for different sets of parton densities.
The `error bars' indicate the expected statistical accuracy
for the PHENIX experiment. Figure taken from$^6$.
\label{fig:fig3}}
\vspace*{-0.5cm}
\end{figure}
The NLO QCD corrections to the direct (non-fragmentation) processes have been
calculated in\cite{ref:nlophoton} and lead to a much 
reduced factorization scale dependence as compared
to LO estimates. In addition, Monte Carlo codes have been 
developed\cite{ref:mcphoton,ref:frixione}, 
which allow to include various isolation criteria and to study photon-plus-jet
observables. The latter are relevant for $\Delta g$ measurements at 
STAR\cite{ref:rhicreport,ref:matthias}.
Since present comparisons between experiment and theory are not fully satisfactory in the
unpolarized case, in particular in the fixed-target regime, considerable efforts have been 
made to push calculations beyond the NLO of QCD by including resummations of large
logarithms\cite{ref:resum}. It is hence not unlikely that a better understanding of
prompt photon production can be achieved soon.
Figure~\ref{fig:fig3} shows $A_{\mathrm{LL}}^\gamma$ as predicted by a NLO QCD
calculation\cite{ref:frixione} as a function of the photon's transverse momentum. The rapidity
cut $|\eta|\leq 0.35$ matches the acceptance of the PHENIX detector.
The important result is that the expected statistical errors are considerably smaller than
the changes in $A_{\mathrm{LL}}^\gamma$ due to different spin-dependent gluon
densities over a wide range of $p_T$. RHIC should be able to 
probe $\Delta g$ in prompt photon production.

\begin{figure}[th]
\centerline{\epsfxsize=3.05in\epsfbox{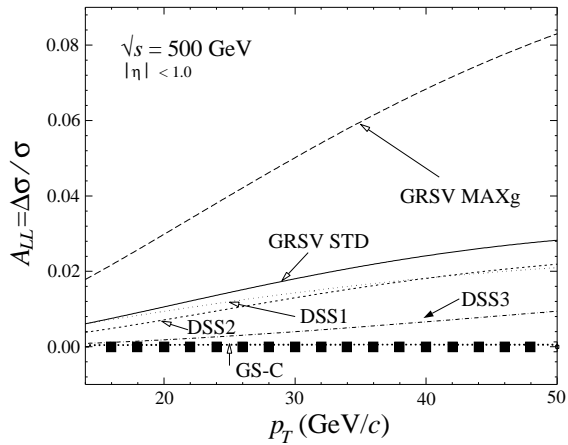}}   
\caption{As in Fig.~\ref{fig:fig3} but now for high-$p_T$
jet production. The `error bars' are for the STAR experiment taking into
account its acceptance. Figure taken from$^8$.
\label{fig:fig4}}
\vspace*{-0.2cm}
\end{figure}
Jets are another key-process to pin down $\Delta g$ at RHIC: they are copiously
produced at $\sqrt{S}=500\,\mathrm{GeV}$, even at high $p_T$,
$15\lesssim p_T \lesssim 50\,\mathrm{GeV}$,
and gluon-induced $gg$ and $qg$ processes are expected to
dominate in accessible kinematical regimes. Due to limitations in the angular
coverage, jet studies will be performed by STAR only. PHENIX can alternatively
look for high-$p_T$ leading hadrons, such as pions, whose production proceeds
through the same partonic subprocesses as jet production. Hadrons have the advantage 
that they can be studied also at $\sqrt{S}=200\,\mathrm{GeV}$ and down to lower 
values in $p_T$ than jets as they do not require the observation of clearly 
structured `clusters' of particles (jets).
On the downside, they require a fragmentation function in the theoretical
description, cf.\ Eq.~(\ref{eq:eq1}), which is, however, fairly well 
constrained by $e^+e^-$ data. 
It should be emphasized that in the unpolarized case, the comparison between
NLO theory predictions with jet production data from the Tevatron is extremely successful.

The NLO QCD corrections to polarized jet production are available as a Monte Carlo
code\cite{ref:nlojets}. Apart from a significant reduction of the scale dependence,
they are also mandatory for realistically matching the procedures used in 
experiment in order to group final-state particles into jets.
For single-inclusive high-$p_T$ hadron production the task of computing the 
NLO corrections has been completed only very recently\cite{ref:nlohadron}.
Figure~\ref{fig:fig4} shows $A_{\mathrm{LL}}$ for single-inclusive jet production at
the NLO level as a function of the jet $p_T$. A cut in rapidity, $|\eta|\leq 1$, 
has been applied in order to match the acceptance of STAR. The asymmetries turn out to be smaller than
for prompt photon production, but thanks to the much higher statistics one can
again easily distinguish between different spin-dependent gluon densities.
Very similar results are obtained for single-inclusive pion production\cite{ref:nlohadron}.

\begin{figure}[th]
\centerline{\epsfxsize=3.0in\epsfbox{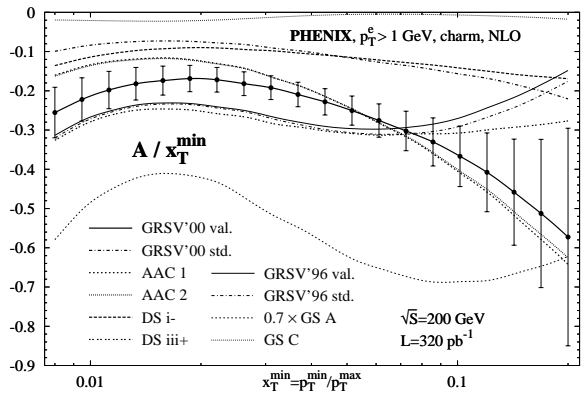}}   
\caption{NLO single-inclusive charm production asymmetry  
(rescaled by $1/x_T^{\mathrm{min}}$) as a function of 
$x_T^{\mathrm{min}}\equiv p_T^{\mathrm{min}}/p_T^{\mathrm{max}}$
for different sets of parton densities. The `error bars' are 
for the PHENIX experiment and include a detection efficiency for 
the channel $c\to eX$ as modeled by PYTHIA. Figure taken from$^{10}$.
\label{fig:fig5}}
\vspace*{-0.2cm}
\end{figure}
The last process which exhibits a strong sensitivity to $\Delta g$ is heavy flavor production.
Here, the LO gluon-gluon fusion mechanism, $gg\to Q\bar{Q}$, dominates unless
$p_T$ becomes rather large.
Unpolarized calculations have revealed that NLO QCD corrections are mandatory for a meaningful
quantitative analysis. In the polarized case they have been completed only very 
recently for single-inclusive heavy quark production\cite{ref:nlohq}. Again, one observes 
a strongly reduced scale dependence for charm and bottom production at RHIC energies.
It turns out that the major theoretical uncertainty stems from the unknown 
precise values for the heavy quark masses\cite{ref:nlohq}.
Since the heavy quark mass already sets a large scale, one can 
perform calculations for small transverse momenta or even for total cross sections
which give access to the gluon density at smaller $x$-values than relevant for jet or
prompt photon production.

Heavy flavors are not observed directly at RHIC but only through their decay 
products. Possible signatures for charm/bottom quarks at PHENIX are 
inclusive-muon or electron tags or $\mu e$-coincidences. The latter provide a much better
$c/b$-separation which is an experimental problem. In addition, lepton detection
at PHENIX is limited to $|y|\leq 0.35$ and $1.2\leq |y|\leq 2.4$ for electrons and
muons, respectively. Since heavy quark decays to leptons
proceed through different channels and have multi-body kinematics, it is
a non-trivial task to relate, e.g., experimentally  observed 
$p_T$-distributions of decay muons to the calculated $p_T$-spectrum of the produced 
heavy quark. One possibility is to model the decay with the help
of standard event generators like PYTHIA\cite{ref:pythia} by computing probabilities
that a heavy quark with a certain $(p_T,y)$ is actually seen within the PHENIX
acceptance for a given decay mode. Figure~\ref{fig:fig5} shows a  
prediction for the charm production
asymmetry $A_{\mathrm{LL}}$ at PHENIX in NLO QCD for the inclusive-electron tag. 
The sensitivity to $\Delta g$ is less pronounced than for the processes discussed
above. It remains to be checked if heavy flavor production at RHIC can be used to 
extend the measurement of $\Delta g$ towards smaller $x$-values.
Also, progress has to be made to solve the long-standing puzzle that the
inclusive $b$-rate as predicted by QCD is too small in unpolarized
$p\bar{p}$, $ep$, and $\gamma\gamma$ collisions\cite{ref:bpuzzle}.

\subsection{Further Information on $\Delta q$ and $\Delta \bar{q}$}
\begin{figure}[th]
\vspace*{-0.3cm}
\centerline{\epsfxsize=0.42\textwidth\epsfbox{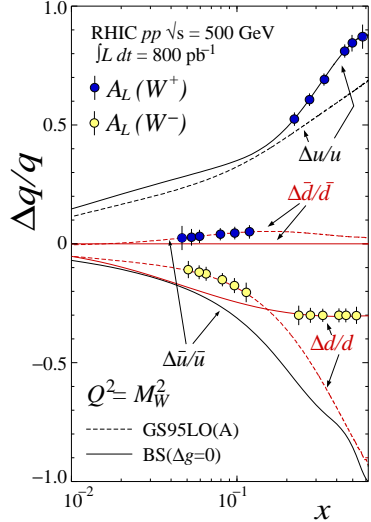}}   
\caption{Expected statistical accuracy for $\Delta q/q$ from
$A_{\mathrm{L}}$ overlayed on two sets of parton densities.
The full [open] circles refer to $A_{\mathrm{L}}(W^+)$
$[A_{\mathrm{L}}(W^-)]$. Figure taken from$^2$. \label{fig:fig6}}
\vspace*{-0.4cm}
\end{figure}

%
Inclusive DIS data only provide information on the sum of quarks and antiquarks
for each flavor, i.e., $\Delta q + \Delta \bar{q}$. At RHIC a separation of
$\Delta u$, $\Delta \bar{u}$, $\Delta d$, and $\Delta \bar{d}$ can be achieved 
by studying $W^\pm$-boson production. 
Exploiting the parity-violating properties of $W^\pm$-bosons,
it is sufficient to measure a single spin asymmetry, $A^W_{\mathrm{L}}$, with only one
of the colliding protons being longitudinally polarized. The idea is to study $A^W_{\mathrm{L}}$
as a function of the rapidity of the $W$, $y_W$, relative to the polarized 
proton\cite{ref:lowboson}.
In LO it is then easy to show\cite{ref:lowboson,ref:rhicreport} 
that for $W^+$-production, $u\bar{d}\to W^+$,
and large and positive (negative) $y_W$, $A^W_{\mathrm{L}}$ is 
sensitive to $\Delta u/u$ ($\Delta \bar{d}/\bar{d}$). Similarly, $W^-$-production probes
$\Delta d/d$ and $\Delta \bar{u}/\bar{u}$. The NLO QCD corrections for $A_{\mathrm{L}}$ 
as well as the factorization scale dependence are small\cite{ref:nlowboson}.
Experimental complications\cite{ref:rhicreport} arise, however, from the fact that 
neither PHENIX nor STAR are hermetic, which considerably complicates the reconstruction of $y_W$.
The anticipated sensitivity of PHENIX on the flavor decomposed quark and antiquark
densities is illustrated in Fig.~\ref{fig:fig6}.

Semi-inclusive DIS measurements, $ep\to HX$, are another probe to 
separate quark and antiquark densities. HERMES has recently published first
preliminary results\cite{ref:beckmann}. The accessible $x$-range for the
$\Delta q$ and $\Delta \bar{q}$ densities is comparable to that of RHIC,
see Fig.~\ref{fig:fig6}, but at scales $Q\simeq 1-2\,\mathrm{GeV}$ rather than $M_W$.
The combination of both measurements can provide an important test of the QCD 
scale evolution for polarized parton densities.

\subsection{Towards a Global Analysis of Upcoming Data}
%
Having available at some point in the near future data on various different reactions, one
needs to tackle the question of how to set up a `global QCD analysis' for spin-dependent
parton densities. The strategy is in principle clear from the unpolarized case:
an ansatz for the densities at some initial scale $\mu_0$, given in terms of some
functional form with a set of free parameters, is evolved to a scale $\mu$ relevant for a
certain data point. A $\chi^2$-value is assigned that represents the quality of the
comparison of the theoretical calculation to the experimental point. The parameters are varied
until eventually a global minimum in $\chi^2$ is reached mutually for all data points.
In practice, this approach is not fully viable since the numerical evaluations of the
cross sections in NLO QCD are usually time-consuming as they require several tedious
integrations. Hence the computing time for a QCD fit easily becomes excessive. 

In the unpolarized case, the wealth of DIS data already provides a pretty good 
knowledge of the parton densities, and
reasonable approximations can be made for the most time-consuming processes. 
For instance, one can absorb all NLO corrections into some pre-calculated `correction
factors' $K$, and simply multiply them in each step of the fit
to the LO approximation for the cross sections which can be evaluated much faster.
In the polarized case, it is in general not at all clear whether such a strategy will work.
Here, parton densities are known with {\em much} less accuracy so far. It is therefore
not possible to use $K$-factors reliably. In addition, spin-dependent parton densities as
well as partonic cross sections may oscillate, i.e., have zeros, in the kinematical
regions of interest such that predictions at LO and the NLO can show marked differences.
Clearly, in the polarized case the goal {\em must} be to find a way of implementing efficiently,
and without approximations, the {\em exact} NLO expressions for all relevant hadronic
cross sections. A very simple and straightforward method based on `double Mellin
transformations' was proposed in\cite{ref:kosower}. Recently, its actual practicability
and usefulness in a global QCD analysis has been demonstrated\cite{ref:global} in a
case study based on fictitious prompt photon data.

\section{Spin Physics at RHIC with Transverse Polarization}
%
At RHIC one can also study collisions of transversely polarized protons\cite{ref:rhicreport}
giving access to the completely unmeasured leading-twist `transversity' 
densities\cite{ref:transreport} $\delta f$.
Upon expressing transversely polarized eigenstates as superpositions of helicity
eigenstates, $\delta f$ reveals its helicity-flip, chiral-odd nature which explains
its elusiveness. Other striking features of $\delta f$ are that no transversity gluon 
density is possible for spin-$1/2$ targets and the fact that 
$\delta f(x,\mu)$ `evolves away' at all $x$ with increasing scale $\mu$.

\subsection{Double Spin Asymmetries $\mathbf{A_{\mathrm{TT}}}$}
\begin{figure}[th]
\vspace*{-0.7cm}\hfill
\begin{minipage}{0.49\textwidth}
\centerline{\epsfxsize=0.95\textwidth \epsfbox{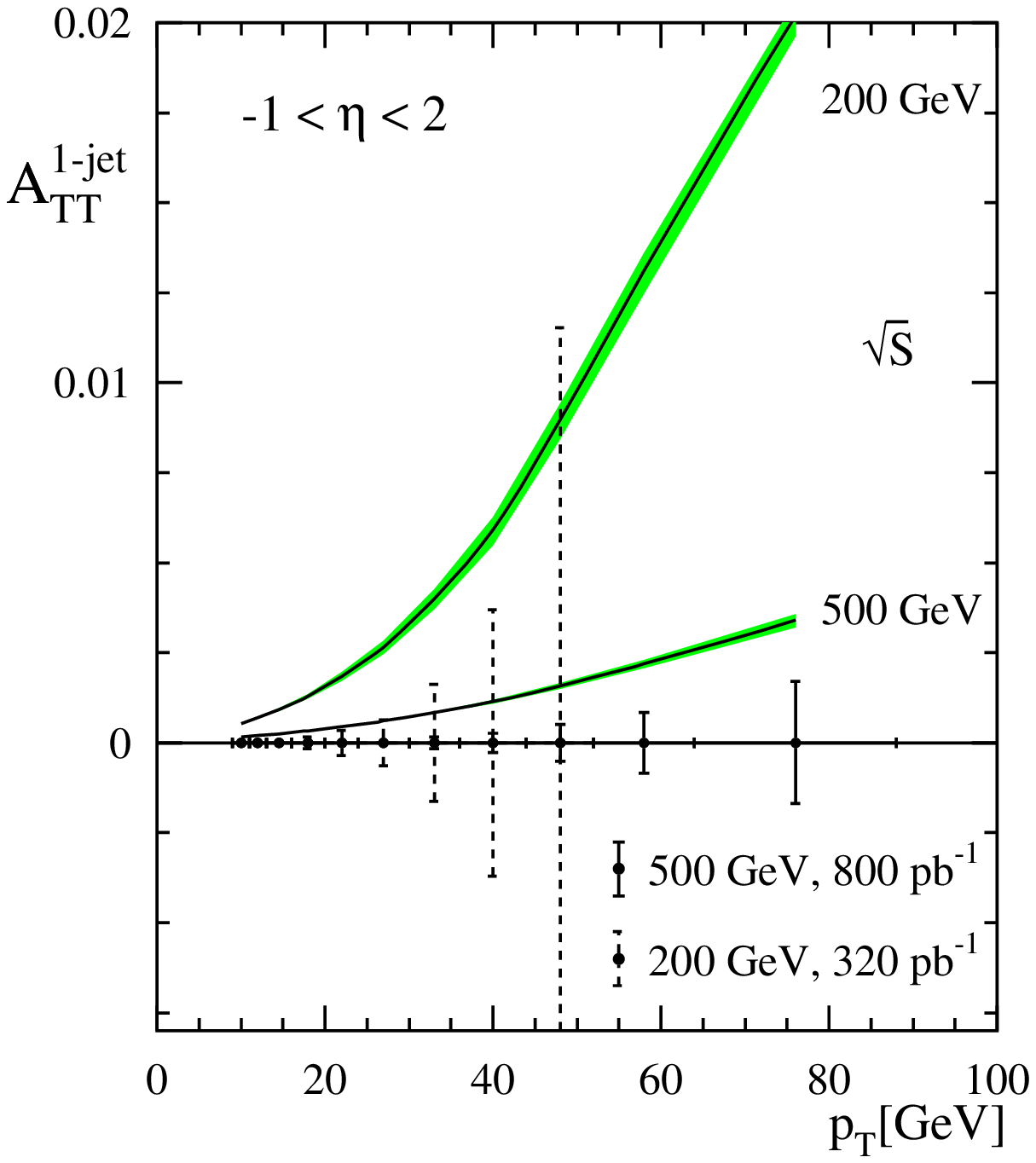}}
\end{minipage}
\begin{minipage}{0.49\textwidth}
\centerline{\epsfxsize=0.95\textwidth \epsfbox{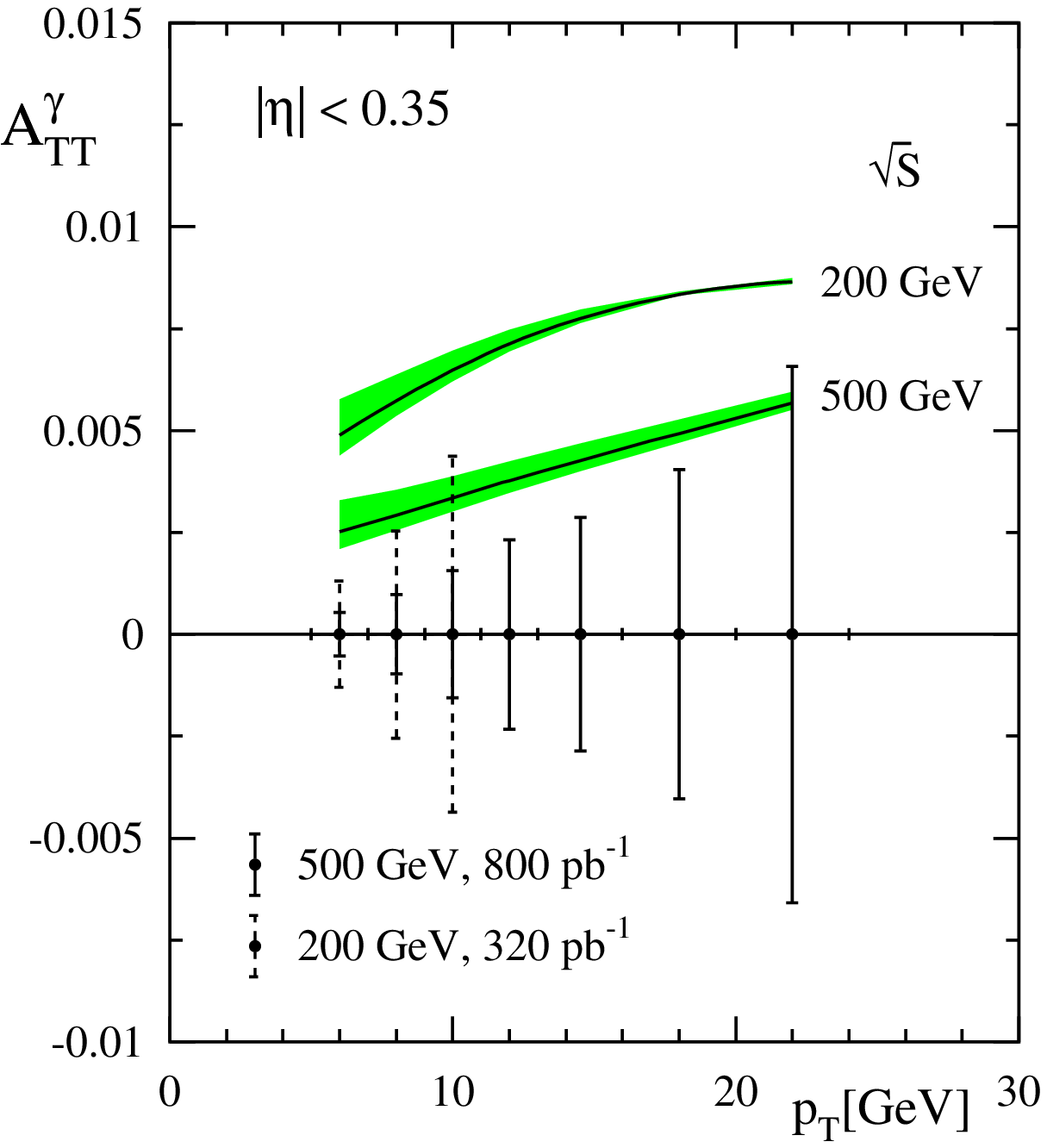}}
\end{minipage}   
\caption{{\bf Left}: upper bound for $A_{\mathrm{TT}}$ for single-inclusive
jet production at RHIC as a function of $p_T$. Jet rapidities are 
integrated over the detector acceptance
$(-1\leq \eta \leq 2)$. The shaded bands represent the uncertainties due
to variations of the factorization scale in the range
$p_T/2\leq \mu \leq 2 p_T$. The expected statistical accuracy is indicated as
`error bars'. {\bf Right}: same as on the l.h.s.\ but now for prompt
photon production. The photon rapidity has been integrated over the range
$|\eta|\leq 0.35$. Figure taken from$^{21}$. \label{fig:fig7}}
\end{figure}

%
The requirement of helicity conservation in hard scattering processes implies that chirality 
has to be flipped twice in order to be sensitive to transversity. One possibility is to
have both colliding protons transversely polarized and to study double spin asymmetries
$A_{\mathrm{TT}}$. Since gluons play an important or even dominant role in almost all
unpolarized production processes, $A_{\mathrm{TT}}$ is expected to be very small 
in general\cite{ref:transreport}.
In addition, $A_{\mathrm{TT}}$ is further diminished by the requirement of a double chirality
flip, which excludes some of the `standard' $2\to 2$ amplitudes to contribute, whereas
the remaining ones are color-suppressed. In principle, the most favorable reaction for determining
transversity is the Drell-Yan process, $pp\to \mu^+\mu^-$, which has no gluonic
contribution in LO. However, a recent NLO study of {\em upper} bounds for $A_{\mathrm{TT}}$,
due to Soffer's inequality\cite{ref:soffer}, $2|\delta f|\leq f+\Delta f$, has revealed 
that the limited muon acceptance for PHENIX threatens to make a measurement of transversity
elusive in this channel\cite{ref:nlody}. Recently is has been shown that, although  $A_{\mathrm{TT}}$
is rather minuscule, jet and prompt photon production can be a useful tool to decipher
transversity at RHIC\cite{ref:att}, see Fig.~\ref{fig:fig7}.
It should be noted that NLO QCD corrections for these processes are still lacking.
Needless to say that such measurements are challenging, albeit not completely impossible.
The experimental finding of a much larger $A_{\mathrm{TT}}$ than theoretically expected would
constitute a new spin puzzle.
%

Alternative observables have only one transversely polarized initial hadron and a 
fragmentation process in the final state that is sensitive to transverse
polarization\cite{ref:transreport}. Several processes have been identified as being
potentially suitable for a measurement of $\delta f$: the production of
transversely polarized $\Lambda$ hyperons\cite{ref:tlambda}, the
asymmetry in the $p_T$-distribution of a hadron in a jet around the jet axis\cite{ref:collins} 
(`Collins effect'), or the interference between $s$- and $p$-waves of a two pion
system\cite{ref:interf} (`interference fragmentation'). Such measurements can be also carried out
at HERMES and COMPASS. However, in all cases involving
these novel fragmentation effects, the analyzing power is a priori unknown and may well
be small, and often there are competing mechanisms for generating the same observable
that do {\em not} involve transversity.

\subsection{Single Spin Asymmetries $\mathbf{A_{\mathrm{N}}}$}
%
\begin{figure}[th]
\vspace*{-0.5cm}
\centerline{\epsfxsize=3.2in\epsfbox{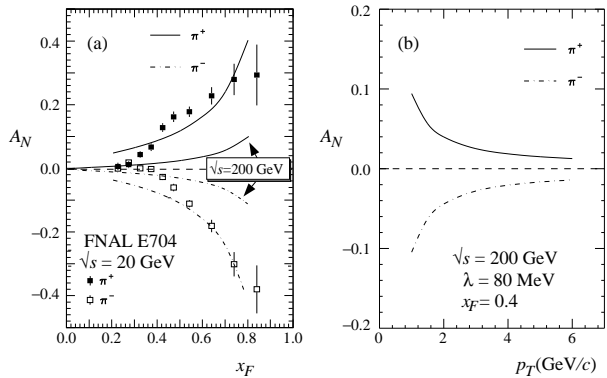}}   
\caption{{\bf (a)}: $A_{\mathrm{N}}$ for pion production in
$pp$ collisions at $\sqrt{S}=20\,\mathrm{GeV}$ compared to 
E704 data. Predictions for RHIC$^{25}$, $\sqrt{S}=200\,\mathrm{GeV}$,
for $p_T=4\,\mathrm{GeV}$ are superimposed.
{\bf (b)}: $p_T$-dependence of $A_{\mathrm{N}}$ for RHIC
at $x_F=0.4$. Figure taken from$^2$. \label{fig:fig8}}
\vspace*{-0.4cm}
\end{figure}
Surprisingly large single transverse spin asymmetries $A_{\mathrm{N}}$ have 
been observed over many years in low-energy fixed-target experiments\cite{ref:transreport},
e.g, for $pp\to \pi X$.
At the leading-twist level for standard ($k_T$-integrated) parton distributions
$A_N$ is exactly zero. One possible explanation is based on a generalized factorization
theorem in perturbative QCD, where a non-vanishing $A_{\mathrm{N}}$ can arise, for instance,
as a convolution of some calculable hard-scattering function with an ordinary
twist-two parton density from the unpolarized proton and a new twist-three quark-gluon
correlation function characterizing the polarized hadron\cite{ref:qiusterman}. 
In a simple model for these correlation functions, which are believed to give the 
dominant contribution to $A_{\mathrm{N}}$ in this approach, 
a qualitative description of the available data 
is possible\cite{ref:qiusterman}, see Fig.~\ref{fig:fig8}, and various predictions have been 
made which can be tested at RHIC. In particular, at RHIC one should see the fall-off 
of $A_{\mathrm{N}}$ with $p_T$ associated with its higher-twist nature, see Fig.~\ref{fig:fig8}.

An alternative approach to  $A_{\mathrm{N}}$ introduces intrinsic transverse momentum
$k_T$ into distribution and fragmentation functions\cite{ref:transreport}.
This opens a Pandora's box with many new and unknown functions. Each of 
three possible mechanisms on its own can account for $A_{\mathrm{N}}$ 
data\cite{ref:transreport}. Needless to 
say that it is very difficult to disentangle all these effects. 
RHIC, with the help of other experiments, can help to shed some light on the 
origin of transverse single spin asymmetries.

\section{Exploring Physics Beyond the Standard Model}
%
Spin observables are also an interesting tool to uncover important new physics.
One idea is to study single spin asymmetries $A_L$ for large-$p_T$ jets. In the 
standard model $A_L$ can be only non-zero for parity-violating interactions,
i.e., QCD-electroweak interference contributions, which are fairly small.
The existence of new parity-violating interactions could lead to sizable
modifications\cite{ref:virey} of $A_L$. Possible candidates 
are new quark-quark contact interactions, characterized by a compositeness scale
$\Lambda$. RHIC is surprisingly sensitive to quark substructure at the 2~TeV
scale, and is competitive with the Tevatron despite the much lower 
c.m.s.\ energy\cite{ref:virey}.
Other candidates for new physics are possible new gauge bosons, e.g.,
a leptophobic $Z'$. Of course, high luminosity and precision as well as
a good knowledge of polarized and unpolarized parton
densities and of the standard model `background' are mandatory.
For details, see\cite{ref:rhicreport,ref:virey}.

\section{Summary and Outlook}
%
With first data from RHIC hopefully starting to roll in soon,
we can address many open, long-standing questions in spin physics 
like the longitudinally polarized gluon density or transversity. 
With data from many different processes taken at high energies where perturbative QCD
should be at work, a first global analysis of spin-dependent parton
densities will be possible.
At the end of RHIC we certainly have a much improved knowledge of the
spin structure of the nucleon, and, perhaps, the next `spin surprise'
is just round the corner. Future projects like the EIC\cite{ref:eic}, which is currently
under scrutiny, would help to further deepen our understanding by probing
aspects of spin physics not accessible in hadron-hadron collisions. The
structure function $g_1$ at small $x$ or the spin content of circularly
polarized photons are just two examples.
  
\section*{Acknowledgments}
%
I am grateful to the organizers for inviting me to an interesting and
lively workshop in Charlottesville.


%
\end{document}